\pdfoutput=1

\documentclass{vgtc}                          

\graphicspath{{figures/}{pictures/}{images/}{./}} 

\usepackage[utf8]{inputenc}
\usepackage[T1]{fontenc}
\usepackage{times}                     

\usepackage{tabu}                      
\usepackage{booktabs}                  
\usepackage{lipsum}                    
\usepackage{mwe}                       

\usepackage{mathptmx}                  
\usepackage{amsmath} 
\usepackage{graphicx}
\usepackage{subcaption}
\usepackage{xcolor}
\usepackage{stfloats}

\definecolor{nipa}{rgb}{0.1, 0.1, 0.9}

\onlineid{2251}

\vgtccategory{Research}

\vgtcinsertpkg

\title{Toward Postural State Classification in Immersive VR with Multimodal data and Explainability Analysis}

\author{Nipa Anjum\thanks{e-mail: nipa.anjum@utsa.edu}\\ %
        \scriptsize University of Texas at San Antonio %
\and Md Irfan Pavel\thanks{e-mail: mpavel1@students.kennesaw.edu}\\ %
     \scriptsize Kennesaw State University %
\and Robert Gonzalez Jr\thanks{e-mail: Robert.gonzalez2@my.utsa.edu}\\ %
     \scriptsize University of Texas at San Antonio
\and Kevin Desai\thanks{e-mail:  kevin.desai@utsa.edu}\\ %
     \scriptsize University of Texas at San Antonio %
\and Alberto Cordova\thanks{e-mail: alberto.cordova@utsa.edu}\\ %
     \scriptsize University of Texas at San Antonio 
\and M. Rasel Mahmud\thanks{e-mail: mmahmud2@kennesaw.edu}\\ %
     \scriptsize Kennesaw State University %
\and John Quarles\thanks{e-mail:  John.Quarles@utsa.edu}\\ %
     \scriptsize University of Texas at San Antonio 
     }

\teaser{
  \centering
  \includegraphics[width=\linewidth]{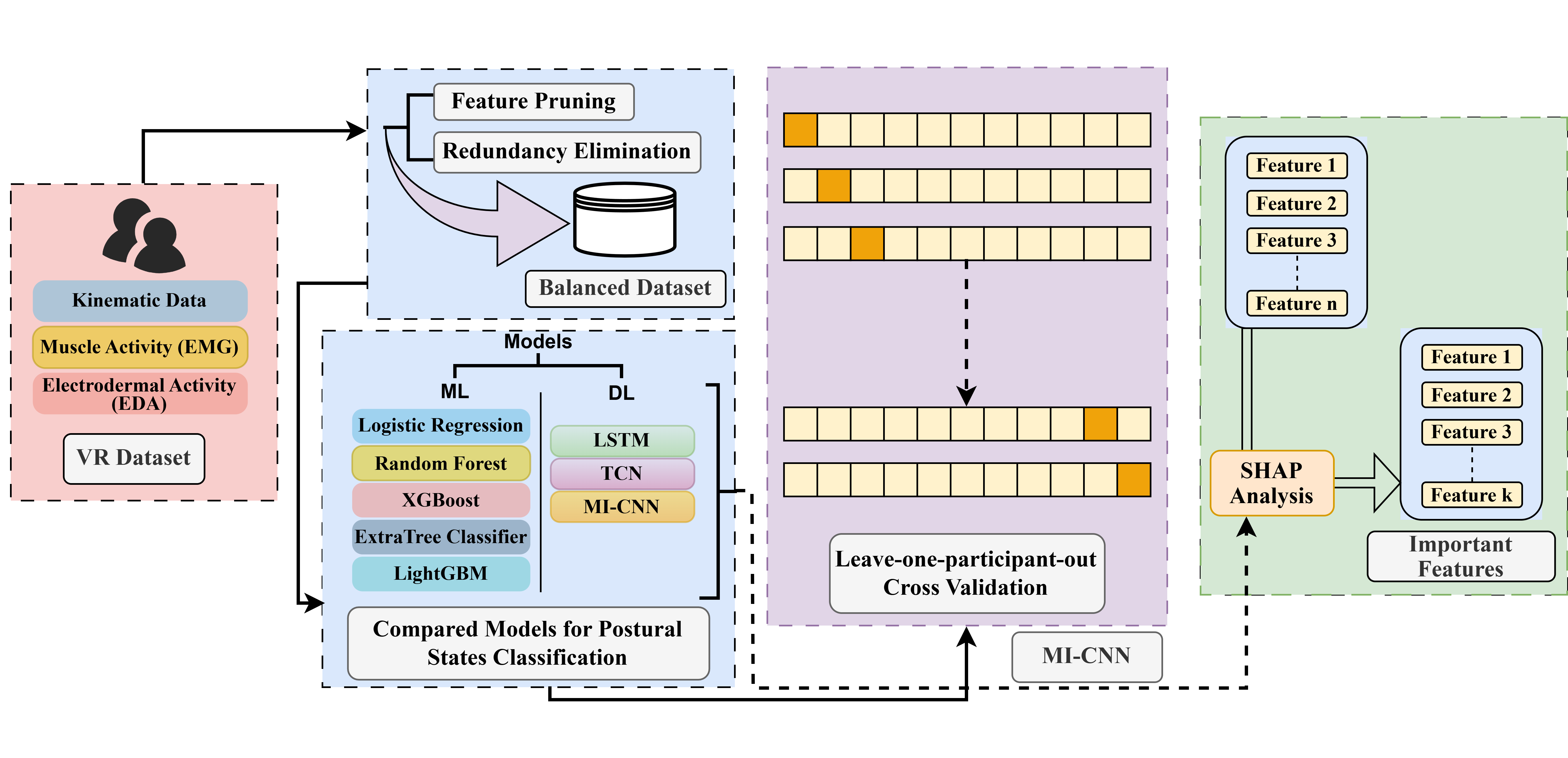}
  \caption{Overview of the study on postural states classification with different models and SHAP interpretability. The pipeline processes VR-collected kinematic, EMG, and EDA signals before training and comparing multiple ML and DL models via leave-one-participant-out cross-validation. SHAP analysis is subsequently applied to the best-performing model, Mamba-inspired CNN (MI-CNN), to identify the most influential features for postural state classification.}
  \label{fig:teaser}
}

\abstract{
Ensuring a safe virtual reality (VR) experience requires effective systems that can predict and respond when users lose their balance. While prior research has focused on predicting falls and motion sickness in VR, most approaches have been regression-based and have not fully explored postural state classification. This study presents an exploratory comparative analysis of machine learning (ML) and deep learning (DL) algorithms for classifying postural states in VR environments affected by visual perturbations. We used a publicly available multimodal dataset that contains kinematic, electromyographic (EMG), and electrodermal activity (EDA) signals. The data were prepared for a binary classification task to distinguish between balanced and imbalanced postural states. Participant-wise downsampling was applied to address class imbalance in the dataset. The ML and DL models that we compared were then evaluated under a Leave-One-Participant-Out (LOPO) cross-validation protocol. Among the models, the Mamba-inspired CNN (MI-CNN) achieved the highest accuracy of 96.76\%. The most important factors influencing the postural state classification were identified using SHapley Additive exPlanations (SHAP) analysis, which enhanced the interpretability of the model. The results from SHAP analysis indicate that kinematic data features are the most dominant in classifying postural states. We also evaluated the MI-CNN model using only the top two-thirds of features ranked by SHAP importance. Despite an approximately 33\% reduction in input dimensionality, the model retained strong classification performance ($0.957 \pm 0.022$ accuracy and $0.957 \pm 0.0218$ F1-score), with only a marginal decrease of approximately 1\% compared to the full-feature model. This study is an initial step toward using multimodal sensing and advanced learning models to better understand balance-related instability in immersive VR environments. The ability to accurately classify an imbalanced postural state may support early awareness of potential fall risks. Also, the findings contribute to the design of safer and more adaptive VR systems. Such systems may respond to early signs of instability and help improve safety and user experience in immersive applications. Code is available at https://github.com/NipaAnjum/MI-CNN.
}

\keywords{Virtual Reality, Balance Prediction, Explainable AI, Machine Learning, Deep Learning, SHAP Analysis, Postural State Classification.}

\begin{document}


\firstsection{Introduction}

\maketitle
Immersive virtual reality (VR) technology is being used for various purposes like games, education, rehabilitation, and training \cite{winther2020design, wang2023asymmetrical, hougaard2022whack, asish2025synthesizing}. While VR offers highly engaging and adaptive environments, it also poses some challenges for human balance and postural stability \cite{mahmud2023auditory}. Sensory conflicts between visual and vestibular inputs, combined with sudden visual perturbations, can induce imbalance or even lead to falls \cite{martinez2018analysing}. Detecting such imbalance events is critical for enhancing user safety and designing adaptive VR systems that respond proactively to early signs of instability.

To investigate balance control under such conditions, recent works have introduced experimental protocols in which participants are immersed in Virtual Environments (VEs) with visual disturbances. Using head-mounted displays (HMDs), participants navigated realistic household or social settings where perturbations are randomly introduced \cite{fang2025social}. Ribeiro et al. \cite{ribeiro2025immersive} developed an immersive VE simulating a household environment, collecting multimodal data, including inertial, EMG, and skin response signals, to analyze compensatory reactions and recovery mechanisms. Similarly, Saini et al. \cite{saini2023novel} introduced a VR paradigm that delivered sudden visuospatial perturbations during overground walking, showing significant changes in gait parameters such as stride length and step width, and highlighting its utility for studying anticipatory and reactive balance control. These studies demonstrate how VR can provide ecologically valid tools for replicating real-world balance challenges. Building on this, previous works have also recorded responses through multimodal sensing, including full-body kinematics with inertial motion capture, muscle activity via electromyography (EMG), and physiological arousal through electrodermal activity (EDA) \cite{wang2021lsl, oishi2021detecting, hartfill2025emg}. Analyzing such high-dimensional multimodal data requires automated computational approaches, making AI models essential for classifying postural states. In this work, postural state classification refers to the binary classification of balanced and perturbation-induced imbalanced states.

Classifying between balanced and imbalanced states in VR is particularly important because it not only enhances user safety but also provides measurable insights into how balance is maintained under challenging conditions. This is highly relevant in rehabilitation and clinical training, where individuals with neurological disorders such as stroke, Parkinson's disease, or multiple sclerosis often face increased balance demands \cite{he2021bibliometric, mariani2023research, hougaard2022whack, pinto2022studying, pruszynska2022towards}. This classification can therefore support the evaluation of rehabilitation progress and clinical outcomes, while also guiding the development of adaptive VR systems that tailor task difficulty and feedback to individual needs.

However, postural state classification in VR has several challenges. First, data can often be highly imbalanced, which can bias models toward the majority class. Second, people exhibit diverse balance control strategies, meaning that models trained on one group may struggle to generalize to new individuals. While AI models frequently achieve competitive results, their limited interpretability remains a barrier to widespread acceptance, particularly in clinical and rehabilitation contexts that demand transparent and explainable decision-making. Overcoming these challenges requires constructing balanced datasets, adopting subject-independent evaluation protocols, and integrating explainability methods that link model outputs to physical and biological factors affecting balance. To address these challenges, we frame our investigation around the following research questions:

\textit{RQ1: How do machine learning (ML) and deep learning (DL) models compare for postural state classification in immersive VR under strict subject-independent evaluation?}

\textit{RQ2: How does performance vary across participants, and what does this reveal about the heterogeneity of balance control in VR?}
\textit{RQ3: How do biomechanical and sensor-level features contribute to postural state classification, and can explainability-guided feature reduction maintain performance?}

To investigate these research questions, we conducted a comparative analysis of classical ML and DL models for postural state classification in VR using a public dataset. To mitigate class imbalance, we applied participant-wise downsampling to construct a dataset with equal numbers of samples for each class. We then used ML and DL models under a Leave-One-Participant-Out (LOPO) cross-validation protocol. Finally, to improve interpretability, we applied SHapley Additive exPlanations (SHAP) \cite{shap} analysis to identify the most influential features that contribute to this classification problem. Rather than focusing only on fall detection after loss of control, this work examines postural state classification to identify destabilizing motion patterns during immersive VR interaction. In VR environments, perturbations can alter postural control before an overt fall occurs, and distinguishing balanced from imbalanced states provides a foundation for adaptive responses such as scene stabilization or difficulty modulation. While the present study does not implement real-time intervention, it demonstrates the feasibility of temporally aware and interpretable imbalance classification under strict subject-wise evaluation.  In conclusion, our work makes the following proposed contributions:
\begin{itemize}
    \item We conducted a comprehensive comparative study of classical ML and DL architectures for VR perturbation induced postural state classification under LOPO evaluation.
    \item We provided a participant-wise performance analysis that explicitly quantifies inter-individual variability, revealing how balance-control heterogeneity influences predictive behavior across unseen users.
    \item We integrate SHAP-based explainability to identify biomechanically meaningful motion features driving classification and demonstrate that performance can be largely preserved under SHAP-guided feature reduction, informing future sensor and feature prioritization.
\end{itemize}
In the following sections we have discussed about prior research works on balance assessment and common methods for prediction, the method of our comparative analysis, and the results.
\section{Related Work}
Research on VR balance assessment, predictive modeling, and explainability has largely progressed separately. Their integration for interpretable postural-state classification in VR remains limited, motivating approaches that both classify imbalance and explain the features driving predictions.
\subsection{VR for Balance Assessment}
VR is increasingly used for balance assessment because it can safely simulate environmental challenges to postural control. Wittstein et al. \cite{wittstein2020use} showed that VR-based sensory organization tests may provide a low-cost alternative to the Equitest system, a clinical computerized dynamic posturography platform widely used for quantitative balance assessment. Wang et al. \cite{wang2021walking} enhanced VR assessments with eye tracking and real-time visualization of body movements, giving useful feedback to clinicians and participants. Altin et al. \cite{altin2020investigation} found that visual attention tasks in VR made balance worse, while auditory tasks helped adaptation, showing how VR can be used to study the link between cognition and balance. Similarly, Dietz et al. \cite{dietz2022walk} explored balance training in VR under stressful environmental conditions, such as height exposure, showing that imitation learning improved balance and reduced stress.

Recent work has also used VR headsets for balance testing. Rosiak et al. \cite{rosiak2024effect} showed that the Meta Quest 2 could track head sway as a low-cost posturography tool, while Sylcott et al. \cite{sylcott2021investigating} found that HTC Vive measurements matched well with force plate data. Wang et al. \cite{wang2021vrgaitanalytics} went further by creating a VR gait system with obstacles, sensory loads, and cognitive tasks, giving real-time feedback. These systems recreate real-world balance challenges in controlled lab settings.

VR has also been used for intervention, showing potential to improve motor function and reduce fall risk through targeted training tasks \cite{wu2025precision, retz2023towards}. Overall, VR provides adaptable, repeatable, and ecologically valid approaches for balance assessment and training, bridging traditional clinical methods with modern immersive technologies.


\subsection{ML and DL-based Approaches for Safety}
Classical ML methods have demonstrated strong performance for balance prediction tasks using sensor data. Chawan et al. \cite{chawan2022person} used smart-floor sensors to assess fall risk in older adults, though class imbalance and overfitting limited robustness. Lin et al. \cite{lin2022automatic} predicted Berg Balance Scale scores using only one thigh sensor and two tasks instead of seven sensors and 17 tasks. Lin et al. \cite{lin2022automatic} improved usability but reported larger errors for participants with BBS scores below 50. Amundsen et al. \cite{amundsen2020gait} introduced SmartFloor for continuous, non-invasive gait monitoring. Askhatova et al. \cite{askhatova2026temporal} combined lower-back IMU signals with handcrafted gait features in a hybrid TCN, while Yu et al. \cite{yu2020novel} used ConvLSTM to classify non-fall, pre-impact fall, and fall stages with low latency. Choi et al. \cite{choi2022deep} used a single waist-mounted IMU and a modified DAG-CNN to distinguish falls, near-falls, and daily activities, achieving over 98\% near-fall accuracy. Overall, these studies show strong potential for sensor-based fall assessment but also highlight limitations in robustness and generalizability.
In parallel, other works have investigated algorithmic strategies for balance prediction. Techniques such as logistic regression, random forests, support vector machines, and gradient boosting have been widely applied. For example, Bao et al. \cite{bao2019automatically} used trunk sway data with an SVM to match physical therapist ratings, outperforming self-assessments but struggling with uneven class distributions. Complementing this, Ta et al. \cite{ta2023exploring} examined whether smaller subsets of the BBS could substitute the full assessment for fall risk prediction. Their findings suggest that reduced task sets may be effective; however, the limited dataset raised concerns of bias and overfitting. Collectively, these studies demonstrate a shift toward simplifying traditional balance assessments while maintaining predictive power; however, they remain constrained mainly to clinical or home-based contexts.

Researchers have also been using techniques for postural control analysis. Choi et al. \cite{choi2024deep} transformed center-of-pressure signals during quiet standing into frequency-domain representations and used CNN backbones such as ResNet-18 to estimate equilibrium scores in patients with dizziness, reporting an absolute difference of about 1.7 between measured and predicted scores. Arias Valdivia et al. \cite{arias2025deep} used recurrent neural networks on force-platform time-series data to classify hemiplegia and diplegia in cerebral palsy, with BiGRU-LSTM models achieving 76.43\% accuracy while also showing differences in postural stability across visual conditions. These findings support the value of temporal and spectral modeling, although broader validation is needed because of the controlled settings and limited populations. 
Beyond fall risk, a growing body of research has applied ML and DL to improve safety and user experience in immersive environments. For example, Qi et al. \cite{qi2025cpnet} introduced CPNet, a DL model that integrates head and eye-tracking, visual complexity, and exposure duration for cybersickness prediction. In contrast, Jeong et al. \cite{jeong2022leveraging} advanced this direction by proposing a multimodal attention-based model trained on 27 participants. Both studies demonstrated the value of combining multiple input modalities but were limited by the narrowness of the datasets and experimental conditions. In a related domain, Elsharkawy et al. \cite{elsharkawy2024adaptive} investigated motion sickness in in-vehicle VR, showing that interactive VR tasks can enhance posture alignment and reduce symptoms, underscoring the importance of considering body posture in adaptive VR design.

Complementary research has also looked at improving tracking and interaction for safety and usability. Burova et al. \cite{burova2020utilizing} combined AR simulation in VR with gaze tracking to efficiently and objectively test industrial applications. Li et al. \cite{li2024wheelposer} developed WheelPoser, a pose estimation system specifically designed for wheelchair users using only four IMUs. Extending pose prediction to locomotion, Kim et al. \cite{kim2024gaitway} proposed GaitWay, an LSTM-based system for trajectory prediction in VR, which performed more reliably than gaze-based approaches in visually complex environments. Retz et al. \cite{retz2023towards} further connected safety and training by introducing VR exergames for fall prevention through an interdisciplinary co-creative process, and An et al. \cite{an2024residual} targeted cognitive fatigue detection via gait videos, using a Residual Graph Convolutional Network to capture posture changes as indicators of fatigue.

Prior work has mainly examined fall risk in clinical or home settings and VR safety factors such as cybersickness, gaze, and pose estimation. Comparatively less attention has been given to classifying balanced and imbalanced states in immersive VR, although this may provide useful insight for future safety-oriented systems.

\subsection{Explainability in VR}
Explainability has become a key consideration in fall detection and risk prediction research, where transparent models can improve both clinical trust and practical deployment. Kim et al. \cite{kim2022fall} used SHAP to identify the most important features from inertial signals collected by smartwatches, enabling efficient fall detection with minimal data transmission. Zhang et al. \cite{zhang2023fall} applied ML models to gait and health data to estimate fall risk in older adults, finding gait frequency and variability to be critical indicators. Gillani et al. \cite{gillani2025clinically} advanced this direction by introducing an explainable framework for radar-based fall detection, combining feature importance, surrogate decision trees, and counterfactual analysis to reduce false alarms and improve clinical acceptance. Together, these works show how SHAP and related explainability tools can highlight biomechanically meaningful features and support more reliable fall prediction systems. Researchers have emphasized explainability in cybersickness detection within VR. Kundu et al. \cite{kundu2022truvr} employed explainable boosting machines to identify gameplay and physiological factors such as exposure length, rotation, and heart rate as key drivers of cybersickness. Dissanayake et al. \cite{dissanayake2025vrsense} introduced VRsense, an explainable system that provides developers with actionable insights into game design choices that influence user comfort. These studies demonstrate the importance of interpretable models in enhancing VR usability.

Extending this line of work, applying SHAP analysis to postural state classification in VR can provide researchers with a new perspective on which features most strongly contribute to imbalance event classification. Such insights can not only improve model interpretability but also guide experimental design by identifying what types of sensor data and balance measures should be prioritized when collecting and analyzing VR-based datasets and systems.
\begin{table}[htbp]
\centering
\caption{Participant-wise Label Distribution in the Dataset before windowing.}
\begin{tabular}{lrr}
\hline
\textbf{Participant} & \textbf{Non-perturbed} & \textbf{Perturbed} \\
\hline
Participant1  & 143420 & 43520 \\
Participant2  & 185612 & 34855 \\
Participant3  & 190992 & 33312 \\
Participant4  & 127912 & 24681 \\
Participant5  & 217880 & 44099 \\
Participant6  & 149186 & 18650 \\
Participant7  & 156766 & 30518 \\
Participant8  & 171632 & 25119 \\
Participant9  & 116792 & 20485 \\
Participant10 & 173087 & 38960 \\
Participant11 & 148904 & 33507 \\
Participant12 & 148904 & 33507 \\
\hline
\end{tabular}
\label{tab:participant_counts}
\end{table}

\section{Methodology}
In this research, we used ML and DL for perturbation-induced postural state classification, validated the models through LOPO cross-validation, and performed feature importance analysis using SHAP (Figure \ref{fig:teaser}). All models were trained and evaluated using the same windowing configuration to ensure a fair comparison across architectures. Specifically, fixed-length sliding windows were generated separately for each participant, with each window containing 100 consecutive time samples. At the sampling frequency of 60 Hz, each window represented approximately 1.67s of multimodal data. A stride of 50 time samples, corresponding to approximately 0.83s, was used, resulting in 50\% overlap between adjacent windows. However, windows were generated separately within each participant, and under LOPO, all windows from the held-out participant were excluded from training. So, overlapping windows could not occur across the training and test sets. Each window was assigned a single label based on the majority class within that window. This windowing procedure was kept consistent across all models. For the tree-based models, a set of window-level statistical features was subsequently extracted from the same windows, including mean, standard deviation, minimum, maximum, median, and range, to provide a compact summary of each window for classification. This gave us a fixed-length vector for each window. By doing this, we reduced dimensionality while preserving discriminative properties essential for tree learning \cite{gomaa2023perspective}. This approach improves computational efficiency compared to flattened raw data \cite{deng2013time, middlehurst2024bake}.

The models were trained and evaluated on a high-performance system featuring an AMD Ryzen 9 7950X3D 16-core processor with 128GB RAM and NVIDIA GeForce RTX 4080 Super and NVIDIA L40S GPUs, providing ample computational resources for efficient processing and analysis of large datasets.
\subsection{Dataset}
For this study, we used a publicly available dataset \cite{ribeiro2025immersive} that examines balance-compensatory reactions in immersive VR environments. The dataset contains multimodal recordings from twelve healthy young adults. There were six male and six female participants with a mean age of $25.09 \pm 2.81$ years. The recordings were collected during VR-induced visual perturbations that were designed to provoke postural adjustments. In the original study, trials were categorized according to the type of visual perturbation and the associated fall-related events, such as lateral disturbances, slips, and trips. Data acquisition combined full-body kinematics from an Xsens MVN Awinda inertial motion capture system, electromyography (EMG) from eight Delsys Trigno channels, and physiological responses from a Shimmer GSR unit.
The dataset was collected through an immersive experimental protocol in a VE that simulated everyday household settings. The environment was developed in Unity and included indoor and outdoor residential scenes such as kitchens, bedrooms, corridors, stairs, and outdoor areas. Participants navigated these environments while wearing an HMD. The household-like virtual environment increased contextual realism compared with simplified laboratory tasks. However, the perturbations remained controlled visual stimuli and did not fully represent spontaneous real-world imbalance events such as physical slips or trips.
Participants experienced several types of visual perturbations during the experiment. These perturbations included scene rotations, translations along different directions, object motion events, and height-related scenarios that could trigger vertigo-like sensations. The perturbations appeared unpredictably while participants moved through the virtual environment. This design encouraged natural compensatory responses and realistic postural adjustments.

During preprocessing, rows that contained duplicate entries or inconsistent signal values were removed to ensure data quality. A closer inspection revealed that the data for Participant 11 and Participant 12 were identical. This situation could introduce data leakage because the model might be evaluated on samples that were already observed during training. To avoid this issue, the data from Participant 12 were discarded. Two features were also missing for Participant 3. These features were removed from the dataset. After these steps, recordings from 11 participants remained, with 1001 features (982 from Xsens, 12 from Shimmer, and 7 EMG) per synchronized time sample. Here, a sample denotes one row of the multimodal time series recorded at 60 Hz. The dataset contains multiple perturbation scenarios across trials and locations within the virtual environment. These scenarios vary in direction and intensity of visual disturbance. As a result, the dataset captured a wide range of compensatory motor responses that include postural adjustments, muscle activation patterns, and body motion dynamics. This diversity makes the dataset suitable for studying balance control mechanisms and for developing computational models that distinguish stable and unstable postural states.

Table \ref{tab:participant_counts} shows the number of perturbed and non-perturbed samples per participant in the original dataset which was initially imbalanced. The majority class corresponded to non-perturbed events, while the minority class corresponded to perturbation-associated events. For this study, we simplified the labels. All perturbed trials were grouped into the imbalanced state (class 1), while non-perturbed trials were treated as balanced or stable state (class 0). To address the unequal class distribution, we applied participant-wise downsampling. For each participant, we retained all samples from class 1 and selected an equal number of samples from class 0. This procedure yielded a balanced dataset across participants while preserving inter-individual variability in balance responses.
\subsection{Machine Learning Models}
We selected five classical ML models: Logistic Regression (LR) \cite{nick2007logistic}, Random Forest (RF) \cite{breiman2001random}, XGBoost (XGB) \cite{xgboost}, LightGBM (LGBM) \cite{lgbm}, and ExtraTrees Classifier (XTC) \cite{geurts2006extremely}. These models represent a diverse set of well-established algorithms with complementary strengths for time series classification. Logistic Regression provides a strong linear baseline and interpretability. Random Forest, XGBoost, LightGBM, and Extra Trees are powerful ensemble tree-based methods known for their robustness to feature correlations, ability to model complex nonlinear relationships, and effectiveness on high-dimensional tabular data. These models are widely used in classification tasks due to their scalability, resistance to overfitting, and capacity to handle heterogeneous feature sets. By evaluating this diverse set, we ensure a comprehensive comparison and establish strong baselines for the DL approaches.

\textbf{Logistic Regression} model was implemented using scikit-learn's LogisticRegression class. The model was trained on flattened sliding windows of raw sensor channels. L2 regularization was used with the saga solver, which is well-suited for high-dimensional datasets. The model was trained with a maximum of 1000 iterations to ensure convergence and used all CPU cores.

\textbf{Random Forest} model was implemented using scikit-learn's RandomForestClassifier. Each time series window was flattened into a single feature vector before training. The model was configured with a small ensemble of shallow trees (10 estimators, maximum depth of 3) and strong regularization, including high minimum samples per split and per leaf, along with subsampling of both features (max features=0.5) and samples (max samples=0.5). The default Gini impurity criterion was used for splitting.

\textbf{XGBoost} was implemented using the XGBClassifier from the xgboost library. The model was trained on statistical features extracted from each window to reduce dimensionality. Strong regularization was applied, including high L1 and L2 penalties, subsampling, and shallow trees (e.g., 5 estimators, max depth of 2). The objective function was set to binary logistic regression, and the evaluation metric was log loss. Tree boosting was performed sequentially using gradient descent.

\textbf{LightGBM} was implemented using the LGBMClassifier from the lightgbm library. The model was trained on the same set of statistical features as the other tree-based models. The active configuration uses a small ensemble of shallow trees (estimators=10, max depth=2, leaves=4) with a learning rate of 0.01 and strong regularization, along with subsampling of both rows and features and large child samples to control overfitting.

\textbf{Extra Trees Classifier} was implemented using scikit-learn's ExtraTreesClassifier. The model was trained on the same set of statistical features as the other tree-based models. The active configuration uses a small ensemble of shallow trees (estimators=10, max depth=3) with strong regularization and subsampling to control overfitting.
\subsection{Deep Learning Models}
We used LSTM \cite{hochreiter1997long} and TCN \cite{lea2017temporal} as our DL models because they are specifically designed to capture temporal dependencies in sequential data. LSTM is well-established for modeling long-term dependencies in time series. TCN leverages dilated causal convolutions and residual connections, enabling efficient learning of both short and long-range temporal patterns with high parallelism. We also utilized a Mamba-inspired CNN \cite{dauphin2017language, gu2024mamba} (MI-CNN) model for this time-series classification. This model combines depthwise causal convolutions with a gating mechanism to capture local temporal patterns in multivariate sensor data, offering a lightweight yet effective architecture for postural state classification. By including these architectures, we aimed to comprehensively evaluate recurrent, convolutional, and gated sequence modeling approaches alongside traditional ML methods for time series classification.

\textbf{CNN-LSTM} model integrates convolutional neural networks (CNNs) with Long Short-Term Memory (LSTM) layers. The initial convolutional layers apply multiple 1D filters to extract local temporal features from the input sequences, capturing short-term dependencies and noise-robust patterns. These features are then passed to LSTM layers, which are designed to model long-term temporal dependencies by maintaining a memory of previous time steps. The combination allows the model to learn both spatial and sequential characteristics of the sensor data. The final dense layers perform classification based on the learned representations.

\textbf{Temporal Convolutional Network (TCN)} architecture is based on stacks of residual blocks containing dilated causal convolutions. Dilated convolutions enable the network to have a large receptive field, allowing it to capture long-range dependencies without increasing the number of parameters excessively. Each residual block consists of two convolutional layers with batch normalization, ReLU activation, and dropout for regularization. Residual connections help with gradient flow and training stability. After several stacks of these blocks, a global average pooling layer aggregates the temporal features, followed by dense layers for classification.

\begin{figure}[t]
    \centering
    \includegraphics[width=\columnwidth]{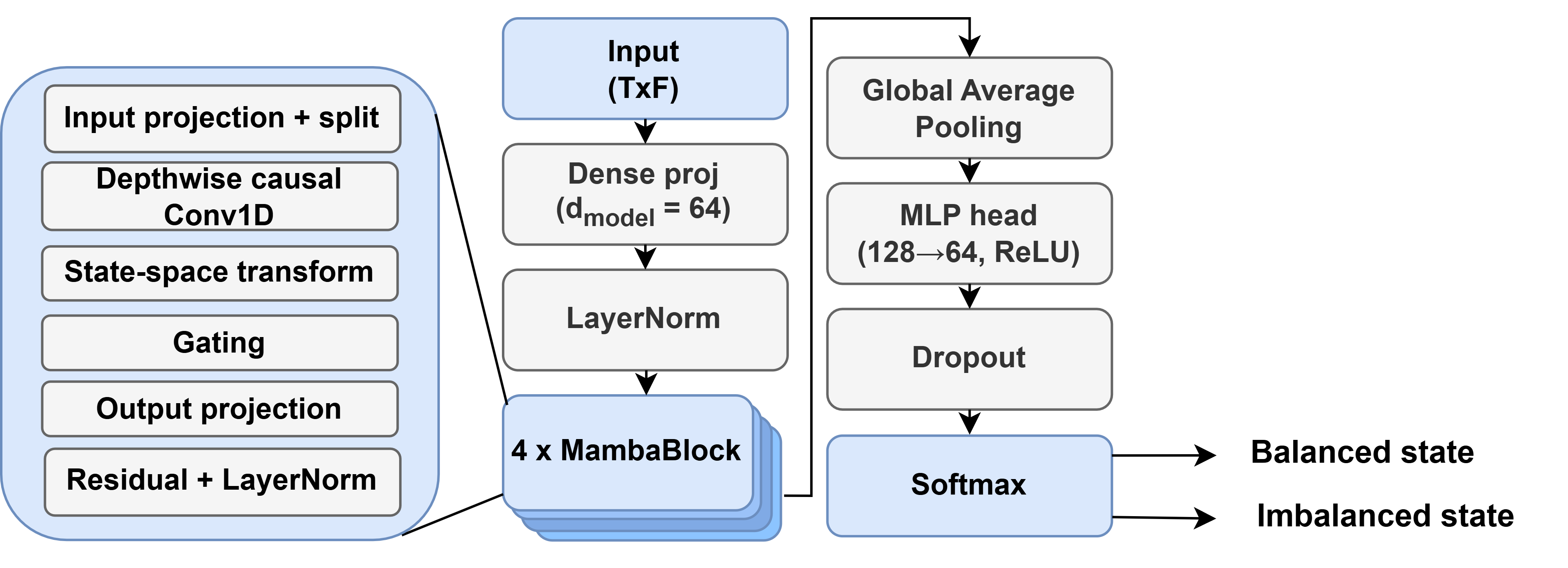}
    \caption{Mamba-Inspired CNN Model Architecture.}
    \label{fig:mi-cnn}
\end{figure}

\textbf{MI-CNN} is a Mamba-inspired convolutional architecture proposed in this work for multivariate time-series classification. Figure \ref{fig:mi-cnn} illustrates the model architecture. While MI-CNN adopts the split-gate-project block structure of Mamba \cite{gu2024mamba}, it replaces the selective state-space recurrence with depthwise causal convolutions and gating mechanism similar to Gated Convolutional Networks (GCN) \cite{dauphin2017language}. This combines Mamba's structural design with the gated linear unit approach of GCNs, resulting in a lightweight TensorFlow-based architecture for time-series classification. Given an input sequence of shape $T \times F$, where $T$ denotes the number of time steps and $F$ denotes the number of input features. Here, $T = 100$ and $F = 1001$, which results in an input window of shape $100 \times 1001$. The model first applies a dense projection to map the input into a latent feature space of dimension $d_{model}=64$. A layer normalization step is then used to stabilize the representation before temporal modeling. The normalized sequence is passed through four stacked Mamba blocks, which form the main sequence encoder of the network.

Each MI-CNN block projects the 64-dimensional representation into two 128-dimensional branches. The temporal branch applies a SiLU activation, followed by a depthwise causal Conv1D layer with a kernel size of 3 and a second SiLU activation. A learned gate, computed using a dense layer followed by Softplus and Sigmoid activations, modulates the convolutional features. The gated output is then multiplied by the SiLU-activated second branch. The resulting features are projected back to 64 dimensions, followed by dropout with a rate of 0.3, a residual connection, and layer normalization. The state-space transform shown in Figure \ref{fig:mi-cnn} represents this simplified state-space-inspired selective gating operation rather than an explicit state-space recurrence. After the four stacked blocks, global average pooling summarizes the temporal representation. The pooled representation is passed through fully connected layers with 128 and 64 units using ReLU activations, each followed by dropout with a rate of 0.3. Finally, a two-unit softmax layer produces the class probabilities. This architecture captures temporal patterns in multimodal sensor sequences for balanced and imbalanced postural-state classification.
\subsection{Experimental Design}
To evaluate generalization across participants, we employed a Leave-One-Participant-Out (LOPO) cross-validation strategy. In each fold, the data from one participant were reserved for testing, while the remaining participants' data were used for training. This ensured that the model was always evaluated on unseen individuals, providing insight into inter-individual variability and the robustness of the classifiers.

We selected LOPO instead of traditional $k$-fold cross-validation (e.g., 5-fold or 10-fold) because our aim was to assess how well the models generalize to entirely new participants. In standard $k$-fold setups, samples from the same participant can appear in both training and testing sets, which risks inflating accuracy and providing overly optimistic estimates of performance. Prior research in related domains supports this concern. Kunjan et al. \cite{kunjan2021necessity} demonstrated that $k$-fold evaluation in EEG-based classification often produces misleading results, whereas a leave-one-subject-out (LOSO) approach provides a more realistic measure of generalization across unseen individuals. Similarly, Esterman et al. \cite{esterman2010avoiding} emphasized that non-independent testing practices can bias results in fMRI analyses and showed that LOSO helps reduce effect size inflation by maintaining independence between training and testing data. Motivated by these findings, we adopted LOPO to obtain a stricter and more reliable evaluation of imbalance prediction models in VR. As a complementary robustness analysis, we evaluated our model using leave-one-perturbation-scenario-out validation, in which all windows from one of the seven perturbation scenarios were excluded from training and used for testing.

Model performance was assessed using accuracy, precision, recall, F1-score, and ROC-AUC, which together capture different aspects of classification quality. Accuracy reflects the overall proportion of correct predictions, precision measures the reliability of positive predictions, recall quantifies the ability to identify actual positives, and F1-score balances precision and recall into a single measure. We report macro-averaged F1-scores to give equal weight to both balanced and imbalanced state classes. ROC-AUC was additionally used to evaluate the models' ability to discriminate between classes. Confusion matrices were also computed to provide detailed insight into class-specific prediction errors.
\begin{table*}[htbp]
\centering
\caption{Overall performance averaged across participants ($mean \pm std$) under LOPO cross-validation.}
\begin{tabular}{l|c|c|c|c|c}
\toprule
\textbf{Model} & \textbf{Accuracy} & \textbf{Precision} & \textbf{Recall} & \textbf{F1} & \textbf{ROC-AUC} \\
\midrule
Logistic Regression & $0.654 \pm 0.097$ & $0.703 \pm 0.088$ & $0.654 \pm 0.097$ & $0.627 \pm 0.127$ & $0.763 \pm 0.093$ \\
Random Forest & $0.797 \pm 0.047$ & $0.811 \pm 0.036$ & $0.797 \pm 0.047$ & $0.794 \pm 0.052$ & $0.866 \pm 0.045$ \\
LightGBM & $0.821 \pm 0.043$ & $0.836 \pm 0.038$ & $0.821 \pm 0.043$ & $0.819 \pm 0.044$ & $0.908 \pm 0.025$ \\
XGBoost & $0.860 \pm 0.030$ & $0.872 \pm 0.021$ & $0.860 \pm 0.029$ & $0.859 \pm 0.031$ & $0.945 \pm 0.014$ \\
Extra Trees & $0.862 \pm 0.028$ & $0.877 \pm 0.021$ & $0.862 \pm 0.028$ & $0.860 \pm 0.028$ & $0.948 \pm 0.012$ \\
CNN-LSTM & $0.935 \pm 0.023$ & $0.937 \pm 0.022$ & $0.935 \pm 0.023$ & $0.935 \pm 0.023$ & $0.984 \pm 0.011$ \\
TCN & $0.935 \pm 0.044$ & $0.937 \pm 0.043$ & $0.935 \pm 0.044$ & $0.935 \pm 0.044$ & $0.979 \pm 0.028$ \\
\textbf{MI-CNN} & $\mathbf{0.968 \pm 0.015}$ & $\mathbf{0.969 \pm 0.013}$ & $\mathbf{0.968 \pm 0.015}$ & $\mathbf{0.968 \pm 0.015}$ & $\mathbf{0.996 \pm 0.003}$ \\
\bottomrule
\end{tabular}
\label{tab:overall}
\end{table*}
\section{Results}
This section reports the results of our experiments, including classification performance under LOPO evaluation and insights from SHAP analysis.
\subsection{LOPO Cross-Validation Performance Across Models}
We evaluated all models using LOPO cross-validation to estimate generalization to unseen individuals. Table \ref{tab:overall} reports the mean and standard deviation of Accuracy, Precision, Recall, F1-score, and ROC-AUC across the 11 held-out folds.
Classical linear modeling yields the lowest performance, with Logistic Regression achieving an F1-score of 0.627 and ROC-AUC of 0.763, indicating that linear decision boundaries are insufficient for separating balanced and imbalanced states in this multimodal VR dataset. Tree-based ensemble methods substantially improved predictive performance. ExtraTrees and XGBoost provided the strongest classical baselines, achieving F1-scores of 0.860 and 0.859 respectively, demonstrating the importance of nonlinear feature interactions across kinematic and physiological channels.
Temporal DL models provided a marked improvement over classical baselines. Both TCN and LSTM achieved F1-scores of 0.935 and ROC-AUC values above 0.97, indicating that explicitly modeling sequential dependencies substantially enhances postural state discrimination. Among all architectures, MI-CNN achieved the highest performance (F1 = $0.968 \pm 0.015$; ROC-AUC = $0.996 \pm 0.003$). Wilcoxon signed-rank tests confirmed that this improvement was statistically significant over all comparison models ($p < 0.01$), including CNN-LSTM and TCN ($p < 0.001$ and $p < 0.01$, respectively).

\begin{figure*}[t]  
    \centering
    \includegraphics[width=0.77\textwidth]{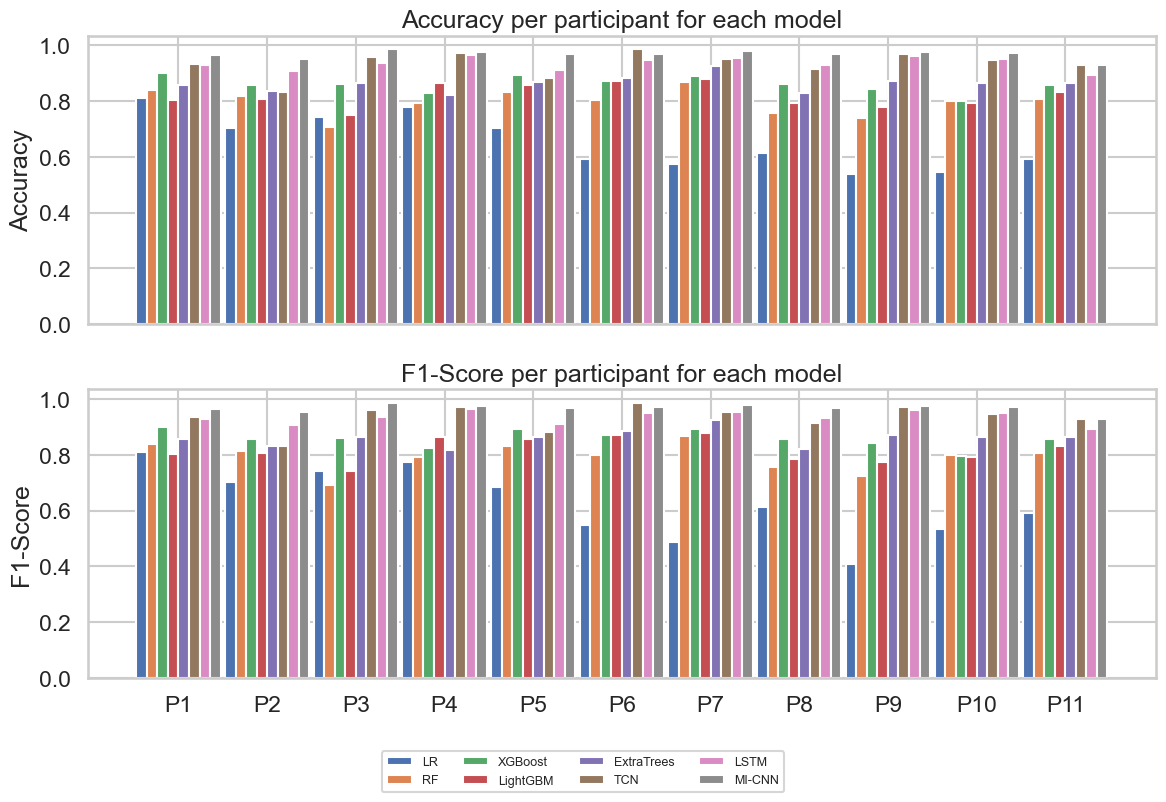}
    \caption{Accuracy and F1-score per participant for each model, showing variability across participants.}
    \label{fig:accf1_per_participant}
\end{figure*}

Because LOPO testing evaluates each model on an entirely unseen participant, performance variation across folds reflects differences in individual balance control strategies. 
Classical models exhibit greater sensitivity to participant-specific patterns, particularly when individual movement strategies deviate from those represented in training folds. Temporal deep models reduced, but did not eliminate this variability. MI-CNN demonstrates the tightest performance range across participants as illustrated in Figure~\ref{fig:accf1_per_participant}, indicating improved robustness to heterogeneous motion patterns. These results suggest that temporal modeling enhances generalization while preserving stability across individuals within the dataset. To further examine whether the LOPO performance was influenced by overlapping windows or repeated perturbation contexts, we conducted two robustness analyses. First, using non-overlapping windows with a stride equal to the 100-sample window length, MI-CNN achieved an accuracy of $0.954 \pm 0.028$, an F1-score of $0.954 \pm 0.029$, and a ROC-AUC of $0.993 \pm 0.006$. Second, under leave-one-scenario-out evaluation, in which each perturbation scenario was excluded from training and used for testing, the model achieved an accuracy of $0.921 \pm 0.038$, an F1-score of $0.928 \pm 0.035$, and a ROC-AUC of $0.977 \pm 0.020$. Performance varied across scenarios, with the fall-from-height/vertigo condition yielding the lowest result (0.855 accuracy, 0.865 F1-score, and 0.941 ROC-AUC), indicating that it was the most challenging unseen scenario. These complementary analyses support the LOPO findings by showing that MI-CNN maintained strong performance when window overlap was removed and when perturbation scenarios were held out, suggesting that the model learned generalizable postural state patterns rather than relying solely on overlapping temporal segments or scenario-specific experimental structure. Furthermore, Pearson correlation analysis between participant demographics (age, height, and weight) and per-participant F1-scores revealed that height was significantly correlated with performance for TCN ($r = 0.861, p<0.001$) and LSTM ($r=0.780, p<0.01$), but not for other models. Participants' weight was significantly correlated only for LSTM ($r=0.656, p<0.05$). No significant associations were found for age. With this small sample size (N=11), these findings are exploratory and require further validation.
\subsection{SHAP Analysis}
To facilitate the interpretability of the MI-CNN model used in this study, SHAP analysis was conducted. SHAP values provide a unified framework for quantifying the contribution of each input feature to the model's predictions. This enables both a global understanding of feature importance across the dataset and a local interpretation of individual predictions. In the context of postural state classification, SHAP analysis allows us to identify which biomechanical features most strongly influence the decision-making process of the model and in what direction.
\begin{figure*}[htbp]
    \centering
    \includegraphics{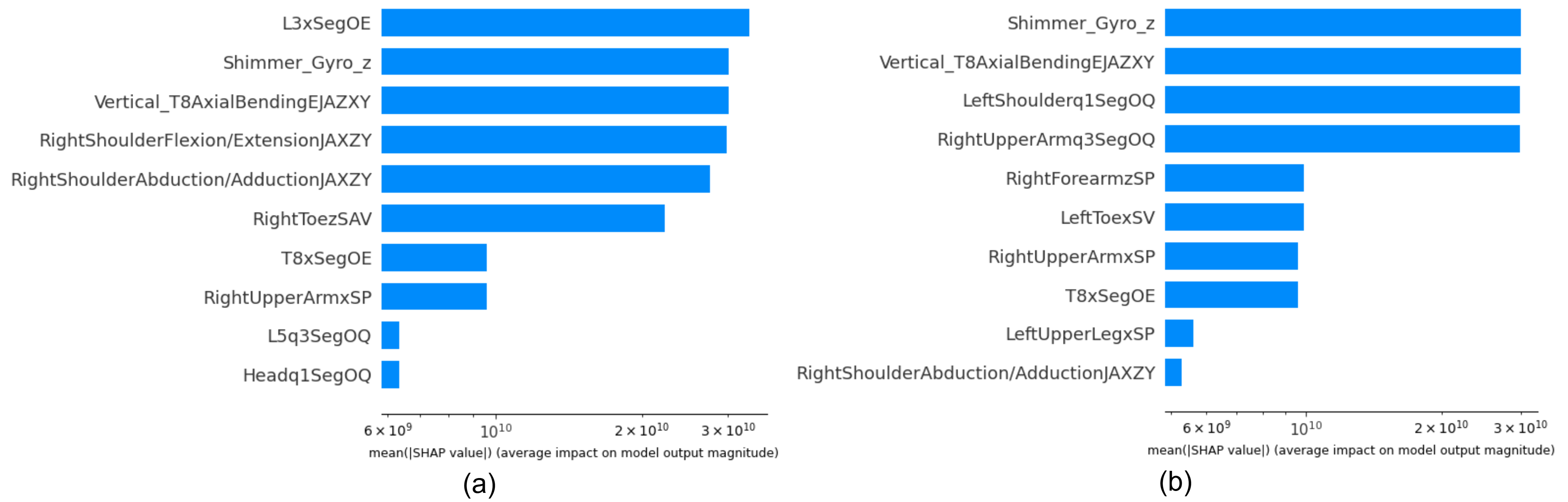}
    \caption{Global SHAP feature importance - (a) Balanced state and (b) Imbalanced state.}
    \label{fig:globalshap}
\end{figure*}
The SHAP explainer was initialized with the MI-CNN model and applied to the test dataset. For each $100 \times 1001$ input window, global importance was computed by averaging absolute SHAP values across time steps and then across held-out windows, while local explanations used signed averages across time.
\subsubsection{Global Explanation}
Global SHAP explanations provide an overall view of which features are most important for the model's predictions by combining information across the entire dataset. SHAP is based on Shapley values from cooperative game theory, which assign each feature a value that reflects its contribution to a single prediction. When these values are calculated for every instance in the dataset and averaged, they produce a ranking of the most influential features used by the model. This approach allows us to see not only which features matter most but also how the model distributes importance across different variables.

For the MI-CNN model, global SHAP explanations capture the role of individual features in shaping complex, non-linear decision boundaries. By examining the average impact of each feature, we can identify those that consistently drive the model toward classifying postural state. Figures \ref{fig:globalshap}(a) and \ref{fig:globalshap}(b) present the SHAP summary plots for the balanced state (class 0) and the imbalanced state (class 1), respectively. The x-axis of each plot represents the mean absolute SHAP value, which indicates the average magnitude of a feature's contribution to the model's output across all samples. Features with higher values have a greater influence on the classifier's decision-making. The features at the top of each bar chart exert the strongest effect on the classifier's output, while those toward the bottom contribute less. This global perspective highlights the dominant biomechanical signals that guide the model's overall decision-making process.

For the balanced state, the most influential features were mainly related to the upper body kinematics and gyroscope signals. On the other hand, for the imbalanced state, lumbar orientation is the most influential feature, followed by trunk bending and gyroscope-related measures. Overall, the global SHAP indicates that trunk motion, lumbar orientation and upper body kinematics are the primary drivers of postural state classification.  Among the most influential features, L3xSegOE captures the forward-backward tilt of lumbar segment, reflecting trunk lean during balance disturbances. Vertical$\_$T8AxialBendingEJAZXY represents lateral bending of the mid-thorax, a key indicator of lateral sway compensation. Shimmer$\_$Gyro$\_$z measures rotational velocity at the forearm, potentially capturing arm swing responses. Right shoulder flexion/extension and abduction/adduction features reflect upper-limb compensatory movements commonly observed during postural recovery.
\begin{figure*}[htbp]
    \centering
    \includegraphics{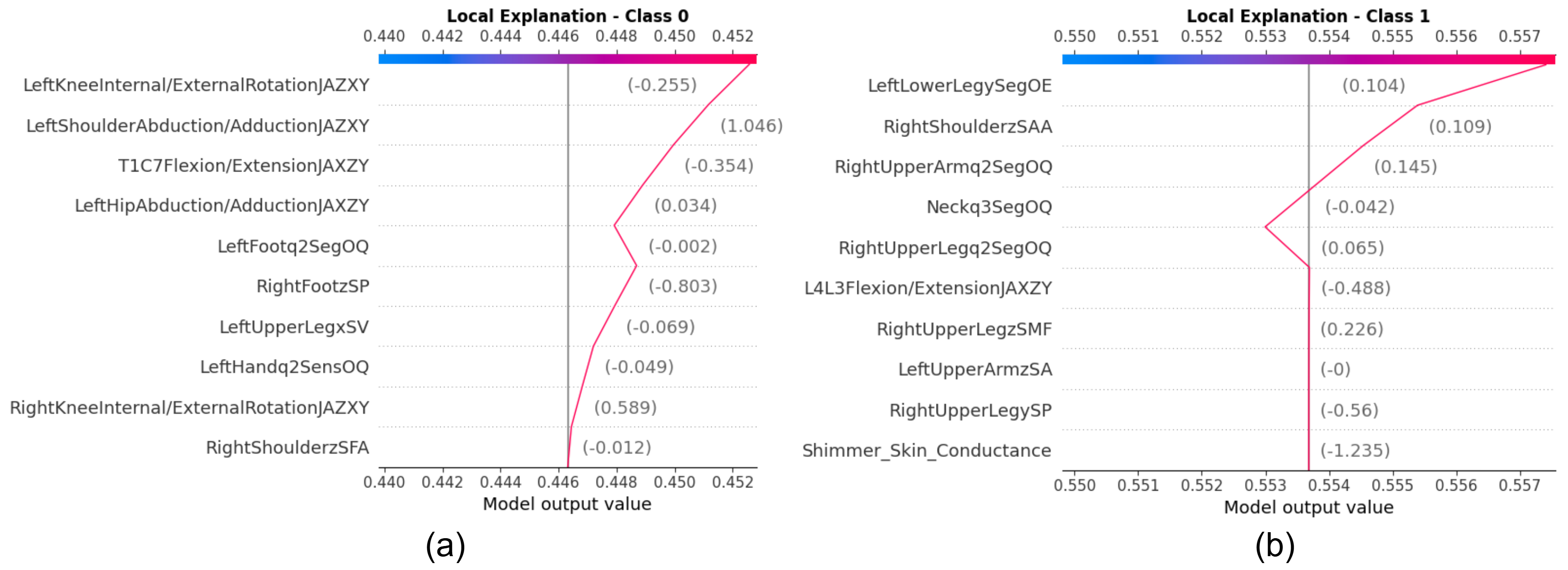}
    \caption{Local explanation result for - (a) Balanced state and (b) Imbalanced state.}
    \label{fig:localshap}
\end{figure*}
\subsubsection{Local Explanation}
While the global analysis provides an overall understanding of the most important features across the dataset, local explanations are crucial for examining how specific predictions are formed for individual samples. In this approach, SHAP values show the contribution of each feature to a single prediction, starting from the model's baseline expectation (the average output across all samples) and moving step by step toward the final decision. This makes it possible to trace which features push a prediction toward the balanced state and which features push it toward the imbalanced state. Unlike global results, local explanations highlight the dynamic roles of each feature, showing that the same feature can have a positive or negative influence depending on the context.

For the MI-CNN model, local SHAP explanations provide insight into how specific features influence individual predictions. The SHAP decision plots begin at the model's base value and show how the displayed features incrementally contribute to the class-specific output. Figures \ref{fig:localshap}(a) and \ref{fig:localshap}(b) show illustrative correctly classified examples for the balanced state (class 0) and the imbalanced state (class 1), respectively, and should not be interpreted as population-level evidence. In each plot, the y-axis lists the features, while the x-axis shows SHAP contributions averaged over time rather than the final softmax probability. The predicted class was determined by the higher of the two softmax outputs.

These illustrative local explanations show that different features can influence individual predictions for the two classes. For the balanced state, knee rotation, shoulder, and foot positioning were the primary drivers. This reflects lower-limb stability and upper-body coordination during balanced posture. For the imbalanced state, lower-leg orientation, lumbar flexion, nech orientation, and skin conductance were more influential. This suggests that the predictions of an imbalanced state are based on a combination of compensatory adjustments of the trunk and lower limb along with physiological responses. 
\subsection{Sensor-Level Contribution}
\begin{figure}[htbp]
    \centering
    \includegraphics[width=\columnwidth, height=4cm]{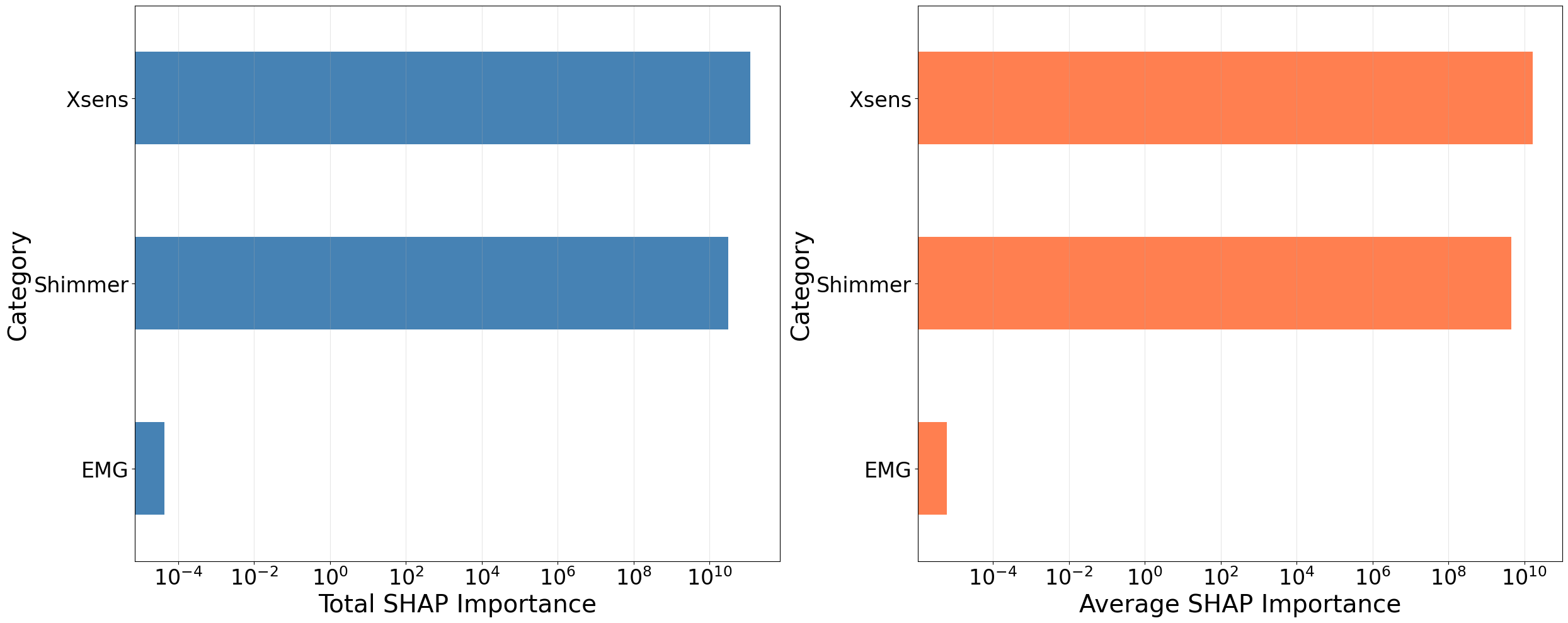}
    \caption{SHAP importance by sensor modality.}
    \label{fig:sensor-level}
\end{figure}

We ensured a fair comparison across modalities by matching the feature count to the smallest sensor features, that is, EMG. Specifically, we selected top-$k$ features per sensor, where $k$ equals the number of EMG features, and aggregated their SHAP values by sensor category to quantify the influence at modality-level. Figure \ref{fig:sensor-level} shows total and average importance across Xsens (full-body kinematics), Shimmer (physiological and inertial signals), and EMG channels.
Full-body kinematic signals contributed the largest share of predictive importance, indicating that segment-level motion dynamics are the primary drivers of imbalance discrimination. Shimmer signals provide a substantial complementary contribution, while EMG channels contribute comparatively less. This suggests that large-scale body motion patterns dominate classification performance, whereas muscle activation features may be redundant or less separable under the current binary grouping.
\subsection{SHAP-Guided Feature Reduction}
To further evaluate the discriminative power of SHAP-identified features, we retrained the model using only the top two-thirds of features. Figure \ref{fig:sensor-level} shows that the Xsens and Shimmer features have the largest modality-level contributions. As a feature-reduction analysis, we retrained the model using the top two-thirds of features from the overall SHAP ranking. The model achieved an average of 95.68\% F1-score and 99.24\% AUC, showing that a reduced feature set can still provide strong performance (Table \ref{tab:before-after-shap}). We also retrained the MI-CNN using the bottom one-third SHAP-ranked features, which achieved 94.16\% F1-score and 99\% AUC. These results suggest that the top two-thirds of features are sufficient to retain most of the full-model performance, but are not strictly necessary, since the bottom one-third alone still supports strong classification performance.
\begin{table}[b]
    \centering
    \caption{Results of MI-CNN before and after SHAP analysis}
    \begin{tabular}{lcc}
        \hline
        Metric & Before SHAP & After SHAP \\
        \hline
        Accuracy & $0.9676 \pm 0.0149$ & $0.9568 \pm 0.0217$ \\
        Precision & $0.9686 \pm 0.0131$ & $0.9590 \pm 0.0189$ \\
        Recall & $0.9676 \pm 0.0149$ & $0.9568 \pm 0.0217$ \\
        F1-Score & $0.9676 \pm 0.0150$ & $0.9568 \pm 0.0218$ \\
        ROC-AUC & $0.9959 \pm 0.0030$ & $0.9924 \pm 0.0080$ \\
        \hline
    \end{tabular}
    \label{tab:before-after-shap}
\end{table}
\section{Discussion}
The comparative evaluation of classical ML and DL models reveals that temporal architectures provide clear advantages for postural state classification in immersive VR. While ensemble methods such as XGBoost and ExtraTrees achieved strong baseline performance, TCN, LSTM, and particularly MI-CNN demonstrated superior results under LOPO evaluation by effectively modeling sequential dependencies in multimodal sensor data. 
The robustness analyses further support the LOPO findings, as MI-CNN maintained strong performance with non-overlapping windows and on unseen perturbation scenarios. Performance was lowest for the fall-from-heights/vertigo scenarios, suggesting that this condition may involve more heterogeneous or scenario-specific postural responses and remains more challenging to generalize. Furthermore, the SHAP-guided feature reduction experiment shows that comparable performance can be maintained using only the top two-thirds of ranked features, suggesting redundancy in the full feature space and supporting the feasibility of more computationally efficient implementations.

The SHAP analysis provides insight into the biomechanical structure underlying model decisions. Global explanations show that trunk, lumbar, and upper-body kinematic features dominate classification, with Shimmer gyroscope signals contributing complementary information, while EMG features show comparatively lower influence. This indicates that trunk and upper-body motion patterns are the primary drivers of postural state discrimination under visual perturbation. Local explanations further show that although trunk-related variables are globally dominant, specific body-segment features (e.g., shoulder or hip orientation) can become more influential in individual predictions. Overall, postural state classification in immersive VR is largely driven by coordinated trunk and upper-body kinematics.

Our findings support prior work showing that deep temporal models are effective for instability-related prediction from sensor signals. Earlier studies showed the value of hybrid TCN \cite{askhatova2026temporal}, ConvLSTM \cite{yu2020novel}, DAG-CNN \cite{choi2022deep}, and recurrent \cite{choi2024deep, arias2025deep} models for fall risk assessment, near-fall detection, and postural control analysis. However, most of these studies were conducted in clinical or home-based settings rather than immersive VR. So, our findings show that the MI-CNN model can achieve very strong performance for perturbation-induced postural state classification in VR under LOPO evaluation. This further supports the idea that capturing local temporal patterns is important for balance-related prediction, while also addressing an underexplored gap in VR safety by focusing specifically on balanced versus imbalanced states in immersive environments. In addition, our SHAP results align with earlier explainability-oriented studies \cite{kim2022fall,zhang2023fall,gillani2025clinically} by showing that interpretable analysis can identify the influential biomechanical signals, with kinematic features emerging as the dominant contributors.

Although the evaluation was conducted offline, MI-CNN is lightweight, with 256,066 trainable parameters and a model size of 3.07 MB. After warm-up, single-window inference required $3.72 \pm 0.46$ ms on GPU and $9.45 \pm 1.21$ ms on CPU. Under the current windowing configuration, a future real-time implementation could process approximately 1.67 s of sensor data per window and update predictions every 0.83 s. These model-level results support the computational plausibility of future online VR integration. End-to-end performance will additionally depend on sensor acquisition, synchronization, preprocessing, and communication with the VR engine.

\section{Limitations and Future Work}
This study provides an initial exploration of postural state classification using ML and DL models and SHAP-based interpretability. The present task focused on classifying balanced and imbalanced states, but considering different types of falls, such as slips, trips, or directional falls, could give a more detailed view of postural control. The dataset contains only eleven healthy young adults and was collected under controlled visual perturbations, so the observed motion patterns may not fully represent spontaneous imbalance in real-world VR. Generalization to older adults, clinical populations, and physical events such as slips, trips, or falls therefore requires further validation. Our exploratory analysis revealed significant correlations between participant height and model performance for some temporal models. However, the dataset is limited to eleven participants. Further investigation with larger participant groups is needed to determine whether factors such as age, height, or weight systematically affect model performance. Although model-level inference latency was measured, the study remains an offline evaluation and does not assess end-to-end sensor streaming, sustained computational load, live adaptation, or user responses to real-time feedback. Future work should evaluate the approach with larger and more diverse populations, more naturalistic imbalance scenarios, and a fully integrated adaptive VR system.

\section{Conclusion}
This study presented a binary classification framework for distinguishing balanced and imbalanced postural states in immersive VR environments using multimodal sensor data. Several ML and DL models were evaluated under an LOPO cross-validation protocol to ensure robustness across individuals. Among them, the MI-CNN model consistently achieved the best overall performance, with statistically significant improvements over other comparison models. Demonstrating the effectiveness of DL models in handling high-dimensional kinematic and physiological features. SHAP analysis further revealed that trunk, lumbar, and upper-body kinematic features were most important for classification. To further evaluate the predictive strength of SHAP-ranked features, the MI-CNN model was retrained using only the top two-thirds of features. The reduced model retained strong performance with only a marginal decrease, showing that fewer key features can still capture most of the important signals and allow for more efficient sensor setups. Overall, this research demonstrates that DL combined with SHAP interpretability offers a powerful framework for understanding and classifying postural states in VR environments. This research also provides guidance for future data collection design, highlighting which sensor modalities and biomechanical features should be prioritized for effective postural state classification. By combining subject-independent evaluation, multimodal temporal modeling, and SHAP-based interpretation, this work provides a methodological foundation for future adaptive VR safety systems. Clinical, rehabilitation, and fall-prevention applications will require further validation with larger and more diverse populations, naturalistic imbalance events, and real-time system-level evaluation.

\section*{ACKNOWLEDGMENTS}
This work was supported by a grant from the National Science Foundation (IIS 2403411). ChatGPT was used during manuscript preparation solely for grammatical review and language polishing.

\bibliographystyle{abbrv-doi}

\bibliography{template}
\end{document}